\documentstyle[aps,prl,preprint]{revtex}

\begin{document}
\draft % \draft command makes pacs numbers print

\title{Wigner crystallization in quantum electron bilayers} 

%\author{G. Goldoni\cite{byline1}$^{1,2}$and F. M. 
%Peeters\cite{byline2}$^2$} 

\author{G. Goldoni$^{1,2}$and F.M. Peeters$^2$} 

\address{$^1$ Istituto Nazionale di Fisica della Materia and 
 Dipartimento di Fisica, Universit\`a di Modena, Via Campi 
213/A, I-41100 Modena, Italy. \\
$^2$ Departement Natuurkunde, Universiteit Antwerpen (UIA), 
Universiteitsplein 1, B-2610 Antwerpen, Belgium}

\date{\today}
\maketitle
\begin{abstract}
The phase diagram of quantum electron bilayers 
in zero magnetic field is
obtained using density functional theory. For large electron 
densities the system is in the liquid phase, while for smaller
densities the liquid may freeze (Wigner crystallization)
into four different crystalline phases;
the lattice symmetry and the critical density depend on the
the inter-layer distance. The phase boundaries between different 
Wigner crystals consist of both first and second order transitions, 
depending on the phases involved, and join the freezing curve at
three different triple points.
\end{abstract}

%      73.20.Dx -- Electron states in low-dimensional structures, 
%      64.70.Dv -- Solid-liquid transitions,
%      64.70.Kb -- Solid-solid transitions 

%\begin{multicols}{2}
%\narrowtext

Coupled electron layers, such as those realized in semiconductor double
quantum wells, are a system of current interest exhibiting a lot of
new physics not present in single layers (SLs). The focus is on the
peculiar properties stemming from the Coulomb interaction between the
layers, such as the recently reported  even denominator fractional
quantum Hall states~\cite{EDFQHE}.   In this paper we are concerned
with the influence of the inter-layer interaction on Wigner
crystallization in electron bilayers (BLs) in zero magnetic field. 

At sufficiently low densities
a degenerate liquid of charges freezes into a
lattice, which is called a Wigner crystal
(WC), as the kinetic energy cost of the localization is more than
compensated by the gain in Coulomb energy. Wigner crystallization
has been investigated intensively over the past few years in 
SLs realized in semiconductor heterostructures;
because a WC forms at low density, disorder-induced localization has
been a serious competing mechanism which has prevented an unambiguous
identification of this state. Normally, high magnetic 
fields are used to quench the
kinetic energy and favour the crystallization, and a SL WC is believed to
form for filling factors $\nu<1/5$~\cite{Goldman}. 

It has been suggested that in {\em bilayer} structures localization 
occurs at higher densities than in SLs, making this an appealing system 
for the investigation of the WC. The reason is that each layer
acts as a polarizable background for the other, favouring the formation
of an inhomogeneous phase~\cite{Swierkowski91}. Evidence of the
formation of a BL WC in high magnetic fields has been found
recently~\cite{Manoharan95} at larger filling factors ($\nu < 1/4$ per 
layer) than in SLs. On the other hand, in contrast to the
SL WC, where the triangular lattice has the lowest
energy~\cite{Bonsall77}, the BL WC might have several
different lattice structures, as a result of the competition between
inter- and intra-layer interactions, leading to a complex phase
diagram~\cite{Falko94,Goldoni,Narasimhan95}. A minimization of
the classical potential energy~\cite{Goldoni} shows that five different
crystal lattices are favoured in different ranges of inter-layer
distance $d$ and charge density $n$. In the classical regime the static
 energy and, therefore, the stable structure depends on the
dimensionless product $d\sqrt{n}$; in contrast, in the quantum regime
the situation is more complicated as, due
to the kinetic energy contribution, $d$ and $n$ do not scale out. 

A comprehensive picture of the phase diagram of the BL WC 
is still lacking. The quantum
Hall regime was investigated recently within the
Hartree~\cite{Narasimhan95} and the Hartree-Fock
approximation~\cite{Zheng95} as a function of the filling factor, the
layer separation, and the inter-layer tunneling. In the quantum
Hall regime several solid phases have been
identified which are analoguous to the classical ones~\cite{Goldoni},
as it is expected from the fact that the high magnetic field quenches the
kinetic energy and leads the system towards the classical regime.
However, the question whether all phases survive quantum 
fluctuations
could not be addressed. Indications of the freezing transition in zero
magnetic field was found within a linear response
theory~\cite{Swierkowski91}, but no information on the stable lattice
structure in the different ranges of $d$ and $n$ was obtained. 

In this paper we calculate the phase diagram of two coupled electron
layers in zero magnetic field by density functional theory
(DFT)~\cite{Lundqvist83}. The successful application of DFT to the
freezing transition in several systems makes this method appealing, in
view of its formal simplicity and the possibility to include accurately
[within the local density approximation (LDA)~\cite{Lundqvist83}]
exchange and correlation contributions to the energy, once these have
been calculated (e.g., from simulations) for the liquid phase. 
DFT allows us to study both homogeneous and
inhomogenous phases on equal footing, and, therefore, to identify both
solid/liquid and solid/solid phase boundaries. The system is modelled
by two layers of mobile electrons with the same average
density~\cite{nota-density}, compensated by a fixed uniform background
of positive charge.  Our results can be summarized as follows: we find
that four different  crystalline phases can freeze from the liquid,
depending on $d$ and $n$. Below the freezing density, structural
transitions between the different lattices occur, the transition being
of the first or second order, depending on the phases involved. Three
triple points separate the liquid and the solid phases. 

To set up a DFT, one has first to identify a
functional of the one-particle charge density $\rho({\bf r})$ which is
a good approximation of the exact free energy of the many-body system 
of interest which, of course, is in general unknown~\cite{functionals}. 
Recently, a functional was proposed for electrons in a SL~\cite{Choudhury95} 
which, in terms of the plasma parameter
$r_s=(\sqrt{\pi n} a_B)^{-1}$ ($a_B$ is the Bohr radius) predicts
freezing at $r_s\sim32$, in good agreement with the value $r_s=37\pm5$
obtained from simulations~\cite{Tanatar89}. This
approach~\cite{Choudhury95} proceeds by writing the density matrix
$\rho({\bf r},{\bf r^\prime})$ in terms of the single-particle density
$\rho({\bf r})$ with the {\em ansatz} 
\begin{equation}
\rho({\bf r},{\bf r^\prime})=f({\bf r},{\bf r}^\prime)
\left[\rho({\bf r})\rho({\bf r}^\prime)\right]^{1/2}.
\label{ansatz}
\end{equation}
The function $f({\bf r},{\bf r}^\prime)$ can be calculated exactly for the 
non-interacting electron liquid, 
$f({\bf r},{\bf r}^\prime)=2 J_1(k_F|{\bf r}-{\bf r}^\prime|)/
k_F|{\bf r}-{\bf r}^\prime|$, 
where $J_1(x)$ is the Bessel function of the first kind and  
$k_F=(2\pi\rho_0)^{1/2}$ is the 2D Fermi wavevector, with $\rho_0$ the 
electron liquid density.
Then  one writes the kinetic energy $E_k$, and the exchange energy $E_x
$ (units of energy are Rydbergs) 
\begin{eqnarray}
E_k = -\int d{\bf r} \left.\nabla_{\bf r} \nabla_{{\bf r}^\prime} 
\rho({\bf r},{\bf r^\prime})\right|_{{\bf r}={\bf r^\prime}} , 
\mbox{~~~~~}
E_x = -(1/2)\int d{\bf r} \int d{\bf r}^\prime |{\bf r}-{\bf r}^\prime|^{-1}
|\rho({\bf r},{\bf r^\prime})|^2 ,
\end{eqnarray}
in the LDA by defining a {\it local} Fermi wavevector 
$k_F({\bf r})=(2\pi\rho({\bf r}))^{1/2}$. 
Inserting (\ref{ansatz}) in $E_k$ and $E_x$ one obtains
the total energy of a SL~\cite{nota-background}
with a (in general inhomogeneous) charge density  $\rho({\bf r})$
\begin{eqnarray}
E_0[\rho] &=& \pi \int d{\bf r}  \left[\rho({\bf r})\right]^{2} +
(1/4)\int d{\bf r} \rho({\bf r})^{-1} 
|{\bf \nabla}_{\bf r} \rho({\bf r})|^2  
+ \int d{\bf r}^\prime \int d{\bf r} |{\bf r}-{\bf r}^\prime|^{-1}
\left[\rho({\bf r})-\rho_0\right] \left[\rho({\bf r}^\prime)-\rho_0\right]
 \nonumber\\
&& -(2/3)\sqrt{2/\pi} 
\int d{\bf r}\left[\rho({\bf r})\right]^{3/2}+
\int d{\bf r} \rho({\bf r}) \epsilon_c(\rho({\bf r})).
\label{E_0}
\end{eqnarray}
 The first two terms stem from $E_k$, the third term is the 
direct intra-layer
Coulomb interaction, with $\rho_0$ the average charge density, and the
last two terms are the exchange and correlation contributions,
respectively. In the correlation term, defined as the difference between the 
total energy and the kinetic and Coulomb contributions,
$\epsilon_c(\rho)$ is the correlation energy of a uniform
liquid of density 
$\rho$ obtained  from quantum Monte Carlo calculations~\cite{Tanatar89}.

The above functional was very successful in predicting the freezing
transition in a SL WC~\cite{Choudhury95}. 
It is tempting to extend this
theory to investigate multi-layered structures. Ideally, one would like
to include the correlation energy for the complete structure, which is
difficult to calculate~\cite{nota-correlazione}. For typical
structures, however, intra-layer correlations are much more important
than inter-layer correlations~\cite{Szymanski94}; the latter,
therefore, can be neglected in a first approximation. We write the
energy functional of the bilayer system as $ E_b[\rho] = E_0[\rho] + E_I[\rho]
$, where 
\begin{equation}
E_I[\rho({\bf r})] =  \int d{\bf r}^\prime \int d{\bf r}
\left[\left|{\bf r}-{\bf r}^\prime\right|^2+d^2\right]^{-1/2} 
\left[\rho^A({\bf r})-\rho_0\right] 
\left[\rho^B({\bf r}^\prime)-\rho_0\right] ,
\label{E_I}
\end{equation}
is the inter-layer Coulomb interaction, and the superscripts A and B
label the charge densities of the two layers. $E_b[n]$ includes (within
LDA) the intra-layer correlation, which mainly determines the freezing
density. In writing $E_b$  we have neglected the tunneling
between the layers as well as the finite thickness of the layers,
taking the charge density profile in the direction normal to the layers
as a delta-function in each layer, which is a reasonable approximation
for $ d/a_B > 1$. 

In DFT one usually solves the Kohn-Sham equations self-consistently in
order to obtain the charge density which minimizes $E_b[\rho]$. A simpler
alternative (from the computational point of view) is to guess a
functional form for $\rho({\bf r})$ with one or more variational
parameters. Therefore, we model the charge density of the electron 
solid as a sum of gaussian distributions~\cite{Senatore90}
centered at the lattice sites ${\bf R}$, 
generated by the primitive vectors ${\bf a}_1$, ${\bf a}_2$, in one 
layer, $\rho^A({\bf r})=(\alpha/\pi)
\sum_{\bf R} \exp(-\alpha |{\bf r}-{\bf R}|^2)$, and at ${\bf R}+{\bf c}$
in the opposite layer, $\rho^B({\bf r})=(\alpha/\pi)
\sum_{\bf R} \exp(-\alpha |{\bf r}-{\bf R}-{\bf c}|^2). $
To minimize the Coulomb energy, $c$ is such that
each site in one layer sits in the centre of a cell in the
opposite layer.  $\alpha$ is the localization parameter; it is zero in
the liquid phase and non-zero for inhomogenous phases, and, for like
densities, it is the same in both layers. 

We have minimized the energy $E_b[\rho]$ with respect to the
localization parameter $\alpha$ and to the lattice structure for fixed
density and inter-layer distance~\cite{nota-technicalities}. Of the
several possible lattices, those listed in Table I
are stable in different ranges of $d/a_B$ and $n$. The resulting phase
diagram is summarized in Fig.~\ref{phd}. First, we focus on the
freezing transition. For large $d$, i.e., weak inter-layer coupling,
the freezing  density tends to the single layer result $r_s\sim 32$. In
this limit, the BL WC consists of two nearly uncoupled triangular
lattices. The two lattices are staggered to minimize the residual
long-range Coulomb interaction (phase D). With decreasing $d$ the
formation of an inhomogeneous phase is favoured by a gain in the
inter-layer interaction and, accordingly, freezing takes place at
larger densities, i.e., smaller $r_s$. If $d/a_B \lesssim 32$,
inter-layer interactions become more important and the liquid freezes
into a lattice with a larger coordination number relative to sites
sitting in opposite layers,  namely two staggered squares (phase B) or
rhombic (phase C) lattices. Along the freezing line of phase C, the 
angle $\theta=\arccos \hat{{\bf a}}_1\cdot\hat{{\bf a}}_2$ 
grows continuosly up to $90^\circ$ with decreasing $d$.
For $d/a_B\lesssim9$, the phase which freezes from the liquid is phase
A. It consists of two staggered, rectangular lattices; the aspect ratio
$\beta=|{\bf a_2}|/|{\bf a_1}|$ at freezing is larger than 1 and its
value depends on $d$. For $\beta=1$, phases A and B coincide.

%By sweeping $n$ through the freezing density ($n_f$), $\alpha$ jumps
%from zero (for $n > n_f$) to a finite value (for $n < n_f$) which changes
%little with $d$ or the lattice symmetry ($\alpha r_s^2 \sim 0.8$). 
%We find a more pronounced
%effect of the lattice geometry on the Lindemann's ratio,
%defined as the root mean square fluctuation around a
%lattice site divided by the lattice parameter. Its value
%at melting for phase B ($\simeq 0.44$) is $\sim 20\%$ larger 
%than in phase D ($\simeq 0.37$), while it is practically independent of $d$. 

When $n$ is swept through the freezing density, $n_f$, the value of the
localization parameter $\alpha$ at which the energy minimum is attained
leaps from zero (for $n > n_f$) to a finite value 
(for $n < n_f$) such that $\alpha r_s^2 \sim 0.6$. This 
corresponds to a very delocalized charge density so that the 
Lindemann's ratio $\gamma$, defined as the root mean 
square fluctuation around a lattice
site divided by the lattice parameter, coincides in practice to its 
upper limit \cite{Likos}, corresponding to a uniform density,
which is, e.g., $\gamma=0.373$ for phase D, and  $\gamma=0.408$ for phase B.

In Fig.~\ref{phd} the open circles are the results of \'Swierkowski {\em
et al.}~\cite{Swierkowski91}, who found a singularity in the linear response
function of the BL electron liquid. These points
correspond very well with our calculated BL WC freezing boundary. In
Ref.~\onlinecite{Swierkowski91} the symmetry of the lattice 
which freezes from the liquid was not identified, but the wavevector at
which the singularity occurs is $q/k_F\simeq2.5$. Since $q/k_F=2.50$
for phase B, while $q/k_F=2.69$ for phase D, it seems that the
inhomogeneous phase found in Ref.~\onlinecite{Swierkowski91} is in fact
phase B, again in agreement with our calculation. 

We next consider transitions between the different solid phases. As
seen above, lowering $n$ at large $d$ freezes the liquid into phase D.
If the density is further decreased, i.e., $r_s$ is increased at fixed
$d$, the intra-layer interaction weakens with respect to the
inter-layer interaction. This lowers the energy of phase C which 
finally becomes energetically favoured. Since phases C and D
belong to different symmetry classes this is a first order phase
transition~\cite{Falko94}. In the region of phase C, 
the energy minimum is at $\theta\neq90^\circ$, while $\theta=
90^\circ$ corresponds to a maximum; on reducing $n$ or $d$, the
energy minimum moves towards $\theta=90^\circ$, which finally
becomes a minimum and the lattice makes a continuos transition 
to phase B. Another  structural phase transition occurs  between phases A and
B, at sufficiently low density, when $d/a_B\gtrsim9$. The
transition is a continuous one, where the aspect ratio becomes larger
than 1 at a critical density, which depends on $d$. An
example of the overall evolution of the aspect ratio is shown in
Fig.~\ref{evolution} for $d/a_B=14$; $\beta$ first reaches a maximum
and then decreases. 
When $r_s \gg d/a_B$, however, the present calculation becomes 
less accurate as
inter-layer correlations beyond the static Coulomb interaction, which 
are not important near the freezing transition~\cite{Szymanski94} and
which were neglected here, become more important. Finally, 
the A/B, B/C and
C/D transition curves join the freezing curve at three triple points where
the solid phases and the liquid have the same energy. 

To conclude, we note that, as an alternative to high magnetic
fields, one can use particles
with a small effective Bohr radius, like electrons in silicon
layers~\cite{Pudalov93} or holes~\cite{Shapira95},
to quench the kinetic energy. The range of
densities for the BL WC in zero magnetic field should be within reach in
coupled layers of holes grown along high index crystallographic
directions~\cite{311}, where both high effective masses, to quench the
kinetic energy, and high mobilities, to prevent defect-induced
localization, can be obtained. 

We acknowledge financial support from the HCM network No.
ERBCHRXCT930374, a NATO Collaborative Research Grant, and the Belgian
National Science Foundation. One of us (FMP) acknowledges discussions
with M. Das during the initial stage of this work and the hospitality
of the Australian National University (Canberra). We are grateful to G.
Senatore for enlightening discussions.

\begin{figure}
\caption{Phase diagram of the electron bilayer. Solid curves: first order 
phase transitions. Dashed curves: continuous phase transitions. Solid dots: 
triple points. Open dots: freezing transition deduced from 
Ref.~\protect\onlinecite{Swierkowski91}.}
\label{phd}
\end{figure}
\begin{figure}
\caption{Evolution of the aspect ratio $\beta$ for phase 
A at $d/a_B=14$ obtained by minimization of 
$E_b[n]$ with respect to $\beta$ and $\alpha$. }
\label{evolution}
\end{figure}
%\end{multicols}
%\widetext
\begin{table}
\caption{Lattice parameters of phases A, B, C and D. 
$a$ is the nearest-neighbour distance. For each phase, the primitive vectors 
${\bf a}_1$ and ${\bf a}_2$, 
the inter-lattice displacement ${\bf c}$, the reciprocal lattice 
vectors ${\bf b}_1$ 
and ${\bf b}_2$, and the charge density $n_s$ are indicated. $\beta=|{
\bf a}_2|/|{\bf a}_1|$ and $\theta$ is the angle between
${\bf a}_1$ and ${\bf a}_2$. } 
\label{geometries} 
\begin{tabular}{lcccccc}
Phase & ${\bf a}_1/a$ & ${\bf a}_2/a$ & ${\bf c}$ & 
${\bf b}_1/(2\pi/a)$ & ${\bf b}_2/(2\pi/a)$ & $n_s a^2$ \\\hline
A (Staggered rectangular) & $(1,0)$ & $(0,\beta)$ & 
$({\bf a}_1+{\bf a}_2)/2$ & $(1,0)$& 
$(0,1/\beta)$ & $1/\beta$ \\
B (Staggered square) & $(1,0)$ & $(0,1)$ & 
$({\bf a}_1+{\bf a}_2)/2$ & $(1,0)$& 
$(0,1)$& $1$ \\
C (Staggered rhombic) & $(1,0)$ & $(\cos\theta,\sin\theta)$ &
$({\bf a}_1+{\bf a}_2)/2$ & $(1,-\cot\theta)$&
$(0,\sin^{-1}\theta)$& $\sin^{-1}\theta$ \\
D (Staggered triangular) & $(1,0)$ & $(1/2,\sqrt{3}/2)$ &
$({\bf a}_1+{\bf a}_2)/3$ & $(1,-1/\sqrt{3})$& 
$(0,2/\sqrt{3})$& $2/\sqrt{3}$ 
\end{tabular}
\end{table}


\begin{references}
\bibitem{nota-density} Layers of like densities are realized 
by gate tuning as to preserve symmetric charge distribution~\cite{EDFQHE}.
\bibitem{nota-background} In semiconductor accumulation layers the
positive background sits in a layer at a distance $d_b$ from the electron
layer. This shifts the energy of the system by a constant $2\pi n
e^2/d_b$, corresponding to the electrostatic energy of a plane
capacitor. As we are only interested in energy differences, this
contribution has been neglected.
\end{references}
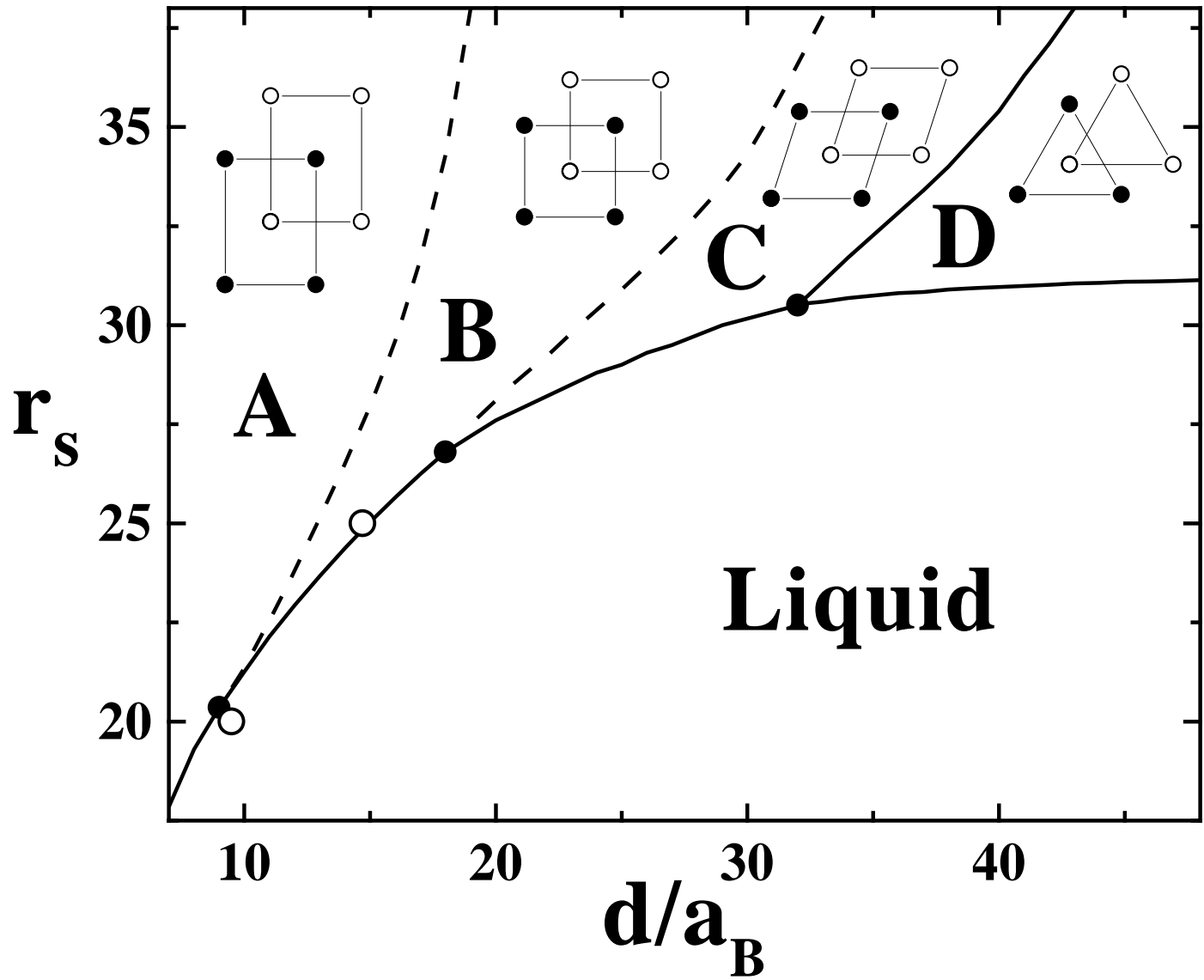
\end{document}